\documentclass[11pt]{article}

\usepackage[english]{babel}

\usepackage{epsfig}
\usepackage{dcolumn}
\usepackage{amsmath}
\usepackage{changebar}
\usepackage{graphicx}
\usepackage{graphicx}
\usepackage{dcolumn} 
\usepackage{bm}      
\usepackage{umlaut}
\begin{document}
\renewcommand{\figurename}{Fig.}
\renewcommand{\tablename}{Tab.}
\title{Black Box QGP
}

\author{ A.~Tawfik\thanks{tawfik@physik.uni-bielefeld.de} \\
 {\small Hiroshima University, 1-7-1 Kagami-yama, Higashi-Hiroshima, 
 Japan }  
}

\date{}
\maketitle

\begin{abstract}
According to extensive ab initio calculations of lattice QCD, the
 very large energy density available in heavy-ion collisions at SPS
 and now at RHIC must be sufficient to generate quark-gluon plasma
 (QGP), a new state of matter in the form of plasma of free quarks and
 gluons. The new state of matter discovered at RHIC seems to be perfect fluid
 rather than free plasma. Its shear viscosity is assumed to be
 almost zero. In this work, I first considered the theoretical and
 phenomenological consequences of this discovery and finally asked
 questions about the nature of phase transition and properties of
 matter. It is important to  
 answer these questions, otherwise QGP will remain a kind of black
 box; one sends a signal via new experiments or simulations or models
 and gets another one from it. I will show that some promising ideas
 have already been suggested a long time ago. I will also suggest a new
 phase diagram with separated deconfinement and freeze-out boundaries and
 a mixed state of thermal quark matter and bubbles of hadron gas. 
\end{abstract}


\section{\label{sec:intr}Introduction}

Theory that describes the dynamics of phase transition to quark-gluon
plasma (QGP) and back to hadronic matter is still failed.
In high-energy experiments, we can only measure produced particles in
the final state, i.e., after chemical and thermal freeze-outs. Consequently,
occurrence of equilibration processes while energy density decreases can
not be confirmed experimentally. Same statement is valid for degrees
of freedom during hadronization processes. It is known that thermal
models work well in the final state. They presume a charge-conserved
hadronic phase and have no access to phase transition. In other words, the
phenomenology of QGP and particularly the dynamics of phase transition
seems to represent a kind of black box.  
This might explain why we used to assume that the deconfinement
temperature $T_c$ at small chemical potential is coincident with the  chemical
freeze-out temperature $T_{fo}$~\cite{TawPT1}. What are the consequences
of this assumption? For example, what type of strongly interacting
matter we have in mind 
that undergoes a phase transition back to confined hadrons, very
much rapidly expands and finally freezes out through chemical processes
without any change in its temperature? Is it possible to describe it
by QCD? Or should we assume that the chemical freeze-out does not exist?

Although the energy density available in heavy-ion collisions at
CERN-SPS~\cite{spsE} and now at BNL-RHIC~\cite{rhicE} exceeds the
critical value calculated in lattice QCD
($\epsilon_c\approx0.7\;$GeV$/$fm$^{-3}$), many physicists still debate
on QGP signatures under these
laboratory conditions. If \hbox{$T_c=T_{fo}$} should hold at small chemical
potentials, phenomenological signals that characterize phase
transition should remain measurable in the final state. 
At least QGP signatures that are not sensitive to a medium, such as 
color screening or $J/\Psi$ dissociation into two leptons have to
survive at \hbox{$T_c=T_{fo}$}. On the basis of current experimental and theoretical
progress, we can think of other solutions, like
$T_c>>T_{fo}$. If $T_c>>T_{fo}$, then QGP
signatures are very much 
contaminated (in-medium-modified) so that the final state does not
reflect QGP production.  Furthermore, we can assume that the
production of QGP requires much higher temperature than the currently
available temperature at RHIC~\cite{rhicT}. 

QCD predicts that matter under extreme conditions is simple but
it does not say anything about matter being ideally formed. For
temperatures larger than $\Lambda_{QCD}$, the matter is in a
deconfined phase. Only a good understanding of the dynamics of phase transition
will help us to characterize QGP.  

\section{\label{sec:nphs}Transition to new phase(s)}

The end of the hadron era was predicted fifty five years
ago~\cite{pomeronchuk51}. No doubt that hadrons will go into a new state
of 
matter at sufficiently high temperatures and densities~\cite{Hag65}. The
wide acceptable  
framework to study the 
phase transition of strongly interacting matter is given by QCD. QCD
predicts that the hot hadronic matter becomes simple. This is not
necessarily weakly interacting. 
QCD has been studied on a lattice for the past thirty
years; QCD Lagrangian has to be 
discretized and everything has to be put on a finite a space-time
lattice. It has been shown that a change in the phase of matter undoubtedly
exists at sufficiently high energy densities. The degree of freedom
markedly increases in a relative narrow region of temperatures. For a
purely gluonic 
system, for which the equation of state can be computed without
approximations, there is a deconfinement phase transition. It is of the
first order 
and the critical temperature is $T_c\sim270\;$MeV~\cite{GlunTherm81}.

\begin{figure}[thb]
\centerline{\includegraphics[width=8.cm]{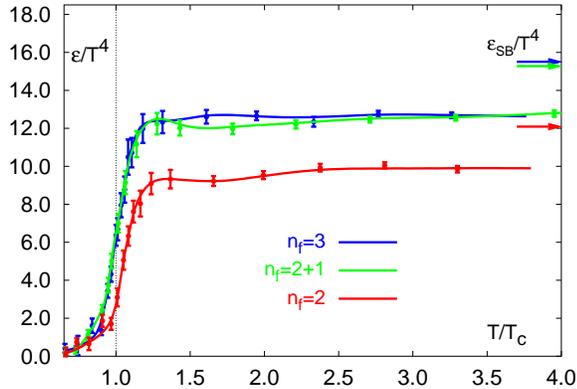}}
\caption{ Energy density in lattice QCD calculated for different
 quark flavors. It is clear that the energy density at the temperature
 $1.2T_c<T>4T_c$ is smaller than the energy density in the
 Stefan-Boltzmann limit.} \label{Fig:1} 
\end{figure}

The difficulties in lattice calculations start when
the fermion sector is switched on. Nevertheless, it has also been observed
that the chiral symmetry is restored at the same critical temperature
$T_c\approx 154-174\;$MeV (depending of quark flavors) as that of
deconfinement transition. The restoration of chiral symmetry
means that the effective mass of quarks forming the 
hadron states becomes zero. Another important consequence of chiral
symmetry breaking  restoration is the disappearance of the mass degeneracy of
hadronic states with the same spin but different parity quantum numbers. Dynamical quarks can only be included in lattice QCD in 
certain approximations. The order of phase transition in full QCD is
not yet completely known. We merely see a rapid change in bulk
thermodynamic quantities. Consequently, the transition is known as
cross-over. The behavior of QCD phase transition at a finite
chemical potential is not yet known from first principle of QCD.
Nevertheless, if we look at the lattice results at a zero chemical
potential, Fig.~\ref{Fig:1}, we find that the 
system at temperatures of $4-5T_c$ remains below the Boltzmann
limit~\cite{Peikert00}; $\epsilon_{SB}\approx g \pi^2 T^4/30$. It seems
that $\epsilon/T^4$ will remain constant at higher temperatures. This
means that the deconfined system is still strongly correlated. This is
also valid at finite chemical potentials.

Apart from the known properties of lattice calculations, like heavy
quark masses and finite lattice size and spacing, we have from this
figure evidence that hadrons undergo a transition to a strongly
correlated phase. Another important finding is the nature of phase
transition. It is not a real phase transition but a rapid change in the
degree of freedom.  Characterizing the nature of matter above $T_c$ is
essential for particle physics. It is important in order to understand
the development of the universe in early stages.

\section{\label{sec:rhic}RHIC results and lattice response}

Using elliptic flow, it was possible to conclude that
RHIC has produced thermalized matter at a very high energy
density~\footnote[1]{The official press release had the title: ''RHIC
scientists serve up perfect liquid''. This mainly has been based on a
comparison with hydrodynamics. Because in the hydrodynamic codes, one
has to introduce 
''many'' inputs, e.g., initial conditions, equation of state, phase
transition, chemical freeze-out and thermal local
equilibration in order to solve the equations of motion
numerically and describe the hydrodynamical evolution,
one has to be very careful with this kind of comparison.}. For 
the first time, hydrodynamics with a zero viscosity can 
describe heavy-ion reactions. In relativistic hydrodynamics, under
the assumption of local thermalization, the number density, current and
energy-momentum tensors are 
\begin{eqnarray}
\delta_{\mu} n_i^{\mu} &=& 0, \hspace*{1cm} \delta_{\mu} J^{\mu}  = 0,
 \hspace*{1cm} \delta_{\mu} T^{\mu\nu} = 0 \nonumber \\
J_i^{\mu} &=& n_i u^{\mu} \nonumber \\
T^{\mu\nu} &=& (\epsilon+P) u^{\mu} u^{\nu} - P g^{\mu\nu} - \eta
 (\delta^{\mu}u^{\nu} + \delta^{\nu} u^{\mu} + \hbox{Tr}) - {\cal
 O}(\zeta) \label{EqHydr1}
\end{eqnarray} 
where $\eta$ and $\zeta$ are the shear and bulk viscosities,
respectively. The number density and entropy density are conserved. 
In strongly interacting quantum fields, the ratio of shear viscosity to
entropy density is as~\cite{Visc1}
\begin{eqnarray}
\frac{\eta}{s} &\approx& \frac{1}{4 \pi}
\end{eqnarray} 
As mentioned above, it was possible to describe RHIC results by
hydrodynamics, even if the last two terms in Eq.~\ref{EqHydr1} are
entirely removed. This means that matter above $T_c$ should also have a zero
viscosity. It is a fluid rather than a free plasma; the
correlations do not completely vanish.

\begin{figure}[thb]
\centerline{\includegraphics[width=10.0cm]{ETAVSS.eps}} \vspace*{-1.2cm}

\centerline{\includegraphics[width=9.cm]{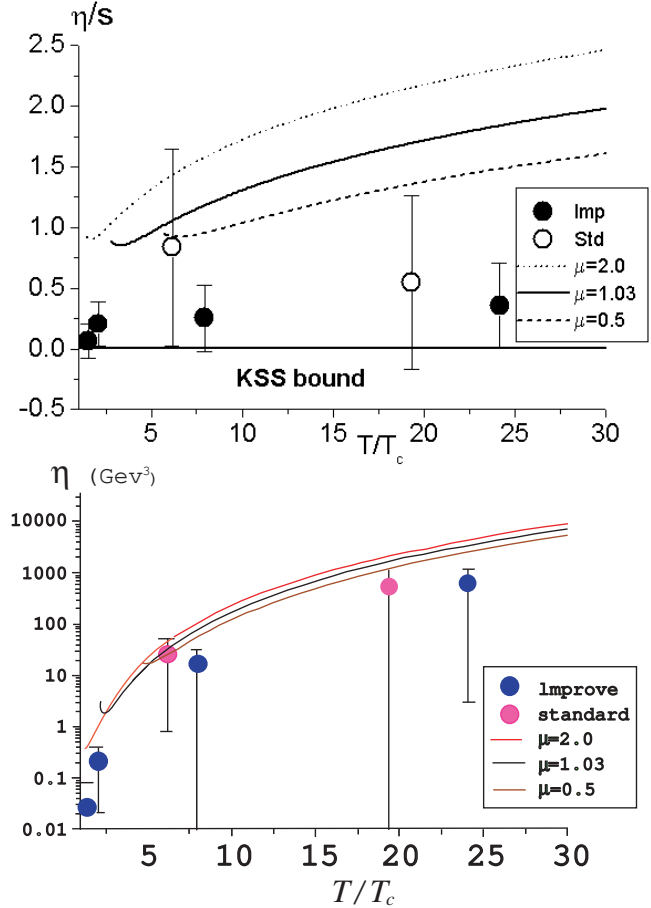}}\vspace*{-.8cm}

\caption{ 
Top: $\eta/s$ obtained by quenched lattice simulations (symbols)
 compared with perturbative results (lines). Bottom: shear viscosity
 in GeV$^3$ units. $\mu$ is a parameter given in Eq.~\ref{Pert}. }
 \label{Fig:Nakam}  
\end{figure}

The pioneering quenched lattice calculations performed by Atsushi~Nakamura of
Hiroshima University for the transport coefficients of 
gluonic plasma~\cite{Nakam1} seem to confirm the above results. The lattice
calculations are performed in the framework of the linear response model. 
\begin{eqnarray}
\eta 
= - \int d^{3}x' \int_{-\infty}^{t} dt_{1} e^{\epsilon(t_{1}-t)}
     \int_{-\infty}^{t_{1}}
     dt'<T_{12}(\vec{x},t)T_{12}(\vec{x'},t')>_{ret}
\label{off-diagonal}
\end{eqnarray}
$<T_{12}(\vec{x},t)T_{12}(\vec{x'},t')>_{ret}$ is related to Green's
function of the energy-momentum tensor. 
%
%
Fig.~\ref{Fig:Nakam} depicts the results for temperatures up to 
$30T_c$. Shear viscosity in the perturbation theory is as~\cite{Arnold} 
\begin{eqnarray}
 \eta = \frac{\eta_1 \cdot T^{3}}{g^{4}(\ln(\mu^{*}/g T)},
 \label{Pert}
\end{eqnarray}
where $\eta_1=27.126$, $\mu^*/T=2.765$, $\mu=T_c/\Lambda_{QCD}$ and $g$
is the running coupling constant.

As mentioned above, only the gluonic sector has been taken into account
in these lattice calculations. The fermionic part entirely vanishes. In this
case, the bulk viscosity is almost zero. It is clear that lattice
calculations agree with the perturbation theory at high
temperatures. However, for $T\leq3T_c$, we find that the lattice results
lie below the perturbative lines. 

It is needed to include quarks. By doing this, we can
describe the transport properties of matter above $T_c$ and make a
better comparison with phenomenological results. Also, we need to
improve the inputs for hydrodynamical codes.\\

\begin{figure}[thb]
\centerline{\includegraphics[width=8.cm]{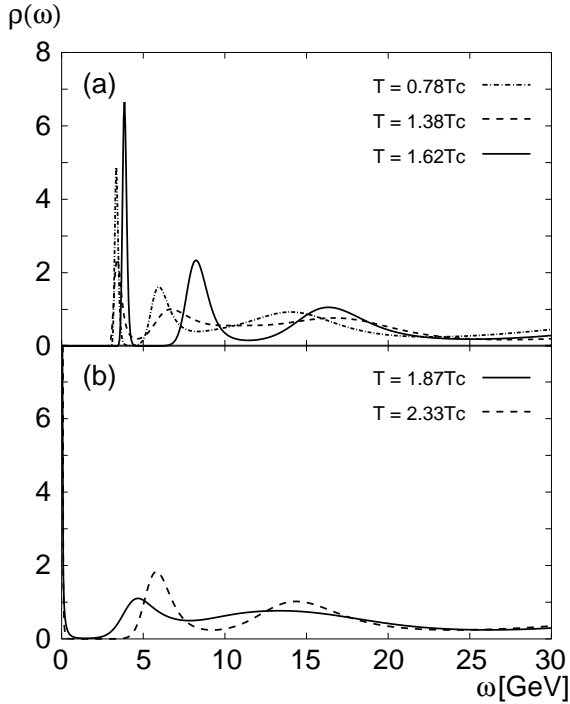}}
\caption{ $J/\Psi$ bound state has been found to survive at almost
 twofold $T_c$. Curves cited from~\cite{JPsi}. } \label{Fig:2}  
\end{figure}

The collision energy is sufficiently high to catapult
hadrons into a new phase. Thus, we now have the situation that the
properties of recently
discovered new state of matter has to be confirmed  
by three independent methods: analytical, numerical and
experimental. 

QCD-like models and/or effective models can be used in the analytical
method. To the author's knowledge, the only analytical method applied
so far -- apart from limitations -- is in~\cite{Visc1}. 

The most powerful numerical method is lattice QCD with dynamical quarks,
which still uses nonphysically heavy or almost vanishing quark masses
and faces a serious challenge when the chemical potential becomes
finite. As mentioned above, the properties of new state of matter are
studied in lattice QCD in  
quenched approximation~\cite{Nakam1}. The lattice results have to be
interpreted with respect to all these approximations. For example, the
critical energy density and temperature at the physical quark masses are
larger than those given above, namely, $T_c\simeq 200\;$MeV,
$\epsilon_c\simeq2\;$GeV$/$fm$^{-3}$. This has the consequence that
$T_c\neq T_{fo}$. The deconfinement point
is not the same as that of chemical freeze-out~\cite{TawPT1}.

The experimental method is the ultimate goal. The proof that a
new phase of matter is produced -- beyond any doubt -- has to be 
adduced experimentally. Experimentalists have maintained this
position since more than two decades. Particle theorists used to tell
them that hadrons for  $T>T_c$  will go into an ''ideal''
deconfined phase and into free  plasma. 

Plasma is a phase of matter in which charges are screened due to the existence
of other mobile charges. This will modify Coulomb's Law. Recent lattice
results on spectral functions show that the $J/\Psi$ bound state can
survive even at $\sim2T_c$, (Fig.~\ref{Fig:2}). In these quenched 
lattice calculations, $T_c\simeq270\;$MeV. As mentioned above, RHIC
results imply that the new state of matter is a fluid rather than 
a free gas. 

On the other hand, when we think about the situation that,
in heavy-ion experiments, one starts and ends up with hadrons, we
realize that the experimental proof is non trivial. Hadrons will be
accelerated, collide against each other and freeze out to  
produce new ''hadrons''. The latter will be counted by 
detectors. This should illustrate the importance of making precise 
theoretical predictions. Nevertheless, I think that experimental
physics is now studied using powerful tools.\\

\section{\label{sec:csrn}Gluon condensates and correlations}

 \begin{figure}[thb]
\begin{center} 
      \epsfig{file=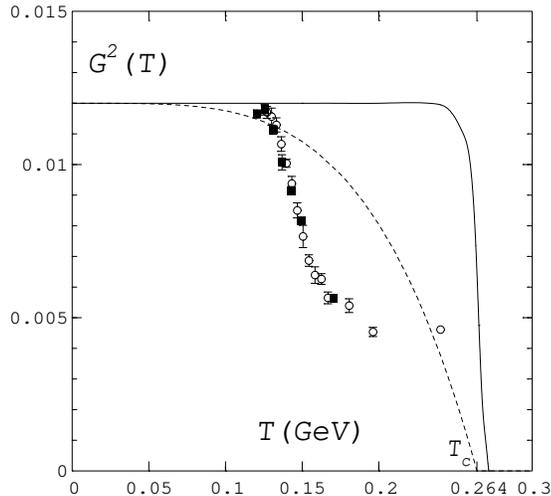,height=70mm,
       bbllx=85,bblly=240,bburx=484,bbury=611} 
\caption{ Gluon condensate of $SU(3)$ (solid line) and an ideal
  gluon gas (broken line) compared with the cases where light (heavy)
  dynamical quarks are included, denoted by open (full) symbols,
  respectively. Curves cited from~\cite{David00} } \label{Fig:3} 
\end{center}
\end{figure}

Many ideas about the complex structure of phase transition have
been suggested. For instance, the soft modes of confined hadrons can
survive the restoration of chiral symmetry breaking~\cite{Hatsuda1} and
that hadron plasma can be formed instead of QGP~\cite{hp}.
Also there are many lines of evidence that deconfined matter don not
behave like free gas. For example, gluon condensates~\cite{shlee1,David00}
remain finite above $T_c$. Gluon condensates have been calculated at a
finite temperature and in the presence of massive
quarks~\cite{David00}. It has been assumed that such quarks due to
strong interactions will modify the thermal properties of gluon
condensates. Therefore gluon condensates remain almost unchanged with
the half of their value in the confined phase~\cite{shlee1}. This means
that matter seems to be sticky~\cite{Brwn2}.   
\begin{equation}
\langle G^2 \rangle_T ~=~ \langle G^2 \rangle_0
~+~m_q\langle\bar{\psi}_q{\psi}_q \rangle_0~
~-~m_q\langle\bar{\psi}_q{\psi}_q\rangle_T~
~-~\langle \Theta^{\mu}_{\mu}\rangle_T.
\label{eq:fullcond}
\end{equation}
Fig.~\ref{Fig:3} depicts this quantity. At $T>T_c$, the condensates at a
zero chemical potential seem to 
be $T-$independent.

Thus, we have concluded that hadronic matter is expected to undergo phase
transition to a new state of matter at high temperatures. The new state
of matter is strongly correlated due to the non vanishing gluon
condensates and the existence of finite hadronic modes.

\section{\label{sec:npd}Phase space and QCD phase diagram}

Phase transition at a very large chemical potential is most
likely of the first order; at $T_c$, only one phase can 
exist. This first-order line is expected to end up with a critical end
point of the second order. Its location was the subject of different lattice
simulations. To date, there has been no final result. In dynamical QCD with
$2+1$ flavors of staggered quarks of physical masses~\cite{Fodor04}, the
endpoint has been localized at $T=162\pm2\;$MeV and
$\mu_b=360\pm40\;$MeV. According to the same reference the critical
temperature at $\mu_b=0$ is $164\pm2\;$MeV. From recent lattice
simulations~\cite{Katz1} and comparisons with the resonance gas
model~\cite{TawRGM}, we assume that $T_c(\mu_b=0)$ should be $\sim
200\;$MeV. The temperature of the chemical freeze-out at a zero chemical
potential is \hbox{$\sim174\;$MeV}.
Another determination of the endpoint has been reported
in~\cite{Shinji1}; $\mu_b\approx420\;$MeV. The corresponding temperature
has not been  estimated.  

We can summarize that the
endpoint is at approximately $\mu_b\approx400\;$MeV or
$\mu_b/T_c\approx2$. This value corresponds to collision energy
$\sqrt{s_{NN}}\approx 9\;$GeV. The lead beam at $40\;$AGeV (SPS) can be used
to scan this region. Note that, at this energy, we have 
fully unexplained peaks in the ratios of strangeness to
non-strangeness particle yields~\cite{TawRt}. That the ratios of
strangeness hyperons to non-strangeness hadrons (pions) all have
a maximum value at the same energy is an indication 
of strangeness asymmetry. This can only be achieved in the plasma
phase~\cite{Reflsk}. i.e., it assumes that the
interacting system at this energy should undergo phase transition to
QGP. The phase transition is likely of the first order. 

On the other hand, we do not observe such a peak at higher energies. It is an
open question, why QGP-signatures e.g., color screening and strangeness
asymmetry are not seen, when colliding energy increases?  It is believed
that, at RHIC and SPS~\cite{Tawold} energies, we have produced the
highest thermalized system ever known. The reason might be the nature of
system produced and the dynamics controlling phase transition. In other
words, the new state of matter might not have the properties of QGP, for
which we have suggested phenomenological signatures. \\

One possibility to explain the difficulties with detecting phenomenological
signatures at low chemical potential is the
non-equilibrium quark occupancy of phase space, i.e., $\gamma_i\neq1$. At
finite temperature $T$, strangeness $\mu_S$ and iso-spin  $\mu_{I_3}$
and baryo-chemical potential $\mu_B$, the pressure of one 
hadron is
\begin{eqnarray}
\label{eq:lnz1} 
p(T,\mu_B,\mu_S,\mu_{I_3}) &=& \frac{g}{2\pi^2}T \int_{0}^{\infty}
           k^2 dk  \ln\left[1 \pm\,
           \gamma\, 
           \lambda_B \lambda_S \lambda_{I_3}
	   e^{\frac{-\varepsilon(k)}{T}}\right], 
\end{eqnarray}
where $\varepsilon(k)=(k^2+m^2)^{1/2}$ is single-particle energy and $\pm$
stands for bosons and fermions, respectively. $g$ is spin-isospin
degeneracy factor. $\lambda=\exp(\mu/T)$ is fugacity parameters. $\mu$
is the chemical potential multiplied by corresponding change. The total
pressure is obtained by summation all hadron resonances. 
$\gamma$ appears in front of Boltzmann exponential,
$\exp(-\varepsilon/T)$. It gives averaged occupancy of phase space relative to
equilibrium limit. Therefore, in equilibrium limit,
$\gamma=1$. Assuming time evolution of system, we can describe  
$\gamma_i$ as ratio between the change in particle number before
and after chemical freeze-out, i.e. $\gamma_i =
n_i(t)/n_i(\infty)$. The chemical freeze-out is defined as time scale,
at which there is no longer particle production and the collisions
is entirely elastic. In case of phase transition, $\gamma_i$ is expected to
be larger than one, because of non-equilibrium processes, large degrees
of freedom and expanding 
phase space.

\begin{figure}[htb] 
\centerline{\includegraphics[width=10.cm]{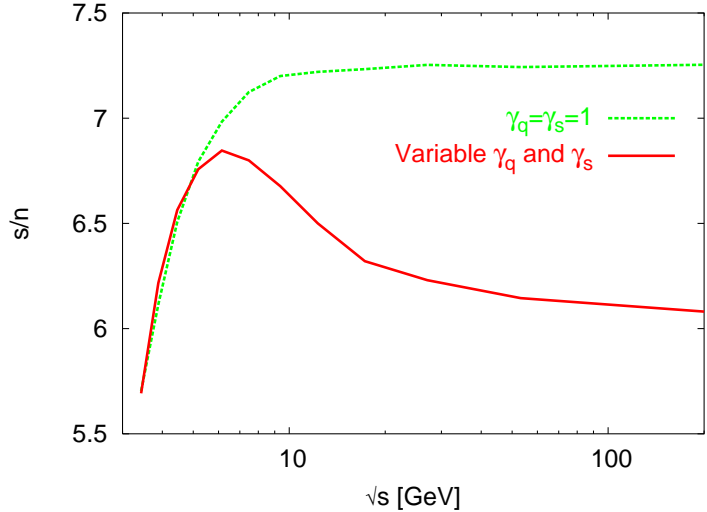}} 
\caption{\label{Fig:5a} 
Entropy per particle $s/n$ as function of $\sqrt{s_{NN}}$. Singularity
 is located at almost the same energy as the peaks of particle ratios of
 strangeness to non-strangeness~\cite{TawRt}.  
}
\end{figure}

The dependence of single-particle entropy on collision energy is related
to averaged phase space density. In Boltzmann limit and for one
particle, we get that
\begin{eqnarray}
\frac{s}{n} &=& \frac{\varepsilon}{T} + 1 - \frac{\mu}{T}
 \label{sOvern} 
\end{eqnarray}
where $\varepsilon$ is single-particle energy. In this expression, $s/n$
directly relates to $\varepsilon$. $s$ is the entropy density. $n$ is
the particle number density. Apparently, $s/n$ gets a maximum
value, when $\mu=\varepsilon$. Because of Boltzmann limit, the
maximum, in this classical system, is unity. Depending on 
$\mu$, we can insert particles into phase space. Maximum occupation is
reached at $\mu=\varepsilon$. Beyond this limit, it is prohibited to
insert more particles. On the other hand, we can expect -- at least
theoretically -- that the occupation value is larger than one, if phase
space itself is changed. The latter situation is most likely provoked by phase
transition. 

The connection between this theoretical discussion and particle ratios of
strangeness hyperons to non-strangeness is given by $\gamma_i$. As in
\cite{TawRt}, $s/n$ plays the same role as $\gamma_i$.

The results of $s/n$ vs. $\sqrt{s_{NN}}$ are depicted in
Fig.~\ref{Fig:5a}. Full grand-canonical statistical set of the
thermodynamic parameters is used. In this case, the complete dependence
of $s/n$ on $T$ and $\mu$ and consequently on $\sqrt{s_{NN}}$, can
straightforwardly be obtained by deriving $s$ and $n$ from Eq.~\ref{eq:lnz1}.

For $\gamma_q=\gamma_s=1$, we find that $s/n$ increases as the energy
raises from AGS until low SPS energies ($\sqrt{s_{NN}}\leq9\;$GeV). For higher
energies, $s/n$ remains constant. This behavior might be an indication
to strong compensation of collision energy in this region. Although,
more energy is introduced to system, number of particles allowed to
occupy phase space remains constant. It is an indirect 
signature that the phase space itself remains constant. \\

For varying $\gamma_q$ and $\gamma_s$, we find a singularity at
$\sqrt{s_{NN}}\approx7\;$GeV. Equivalently, phase space is maximum at this
energy. At higher energies, $s/n$  decreases. Although, energy increases
and consequently the produced particles, single-particle
entropy decreases. This means that phase space shrinks. At RHIC energies,
the shrinking becomes slower than at SPS. If this model 
gives the correct description, we now might have for the first time a
theoretical explanation for dependence of phase space on energy. The
phase space at SPS energy is apparently larger than at RHIC. Same
behavior has been found experimentally~\cite{Gudima}. The consequences
are that QGP might be produced at SPS and detecting its signatures at
RHIC might be non trivial. \\

\begin{figure}[thb] 
\centerline{      \epsfig{file=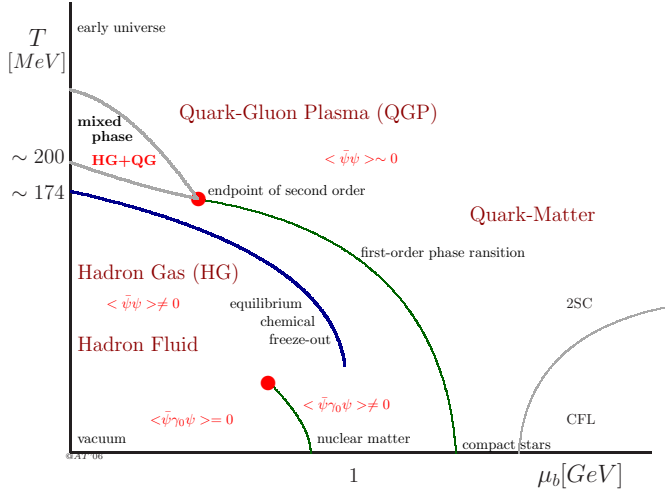,height=70mm,
       bbllx=85,bblly=350,bburx=684,bbury=711} 
}
\caption{ Phase diagram of QCD. The values of the chiral symmetry
 order parameter  $<\bar{\psi}\psi>$ are also given. We distinguish
 between the freeze-out and transition lines. At small chemical
 potentials, the hadrons most likely undergo a smooth transition to a
 thermal phase accommodating a hadron gas and  a fluid of quarks and
 gluons. At hight $T$, the mixed phase entirely disappears. Matter will
 then be formed in QGP. } \label{Fig:my}  
\end{figure}

The situation is different at small chemical potentials. It is most
likely that the transition can not be characterized by a line. As shown
above, the transition in this region is a rapid cross-over. The critical
temperature is an average, (Fig.~\ref{Fig:1}). Therefore, we expect that the
transition is a kind of wide boundary. From Fig.~\ref{Fig:1}, we can
roughly estimate that the transition boundary has a width of at least $4T_c$. 
At temperatures up to $4T_c$, the degrees of freedom in the new phase of
matter are $30\%$ smaller than that in the Stefan-Boltzmann limit,
(Fig.~\ref{Fig:1}).  

There are indications that the hadronic bound states can survive
above $T_c$, (Fig.~\ref{Fig:2}). We have discussed the
hydrodynamical results at RHIC energies. From all these results, it
seems physically consistent to suggest that the new 
phase of matter is a mixture of the surviving hadron gas and a 
fluid of quarks and gluons~\cite{Gyul84}. The hadron gas might form bubbles
inside the quark fluid. This explains two important features, the
degrees of freedom and existing hadronic bound states above $T_c$. \\

Based on the above discussion, we suggest a ''new'' phase diagram in
Fig.~\ref{Fig:my}. We separate the line of equilibrium chemical freeze
out from that of phase transition. At a zero chemical potential, 
phase transition takes place at $T\sim200\;$MeV~\cite{TawPT1,Katz1}. The
equilibrium chemical freeze-out takes place at
$T\simeq174\;$MeV~\cite{TawPT1}. The two lines slightly become shorten with
increasing chemical potential. Phase transition at chemical 
potentials up to the value corresponding to the endpoint is no longer a
cross-over. Consequently, QGP is expected at much higher temperatures
than $T_c$. Between the hadron gas and the QGP there is a mixed phase, in
which the two degrees of freedom partly exist. 
The properties of matter in the mixed phase are as follows. 
\begin{itemize}
\item It is an ordinary ''nearly perfect'' fluid (it is not free gas or
plasma),  
\item It has very small but finite viscosity (it is not a superfluid), 
\item It has electric resistance (it is not a superconductor), 
\item It is a strongly correlated matter, since hadronic bound
	   states can survive above $T_c$. 
\item It is a sticky matter, since gluon condensates remain
	   finite above $T_c$.
\end{itemize}

In the light of these properties, we need to answer the following
questions:  
\begin{itemize}
\item What is the order of phase transition at small chemical
      potentials (related to RHIC, LHC or early stage of universe)? What is the
      order of phase transition from the mixed phase to QGP at very
      high temperatures? 
\item How can we confirm the transition at small chemical potentials 
      phenomenologically? 
\item What are the in-medium modifications of hadron properties at
      finite $T$ and $\mu_b$~\cite{Taw061}? 
\item What are transport properties above $T_c$?
\end{itemize}
We have to answer all these questions before we face a much difficult
challenge with CERN-LHC and GSI-FAIR. We have to have clear theoretical
descriptions and non ambiguous phenomenological signatures describing
the QCD phase diagram. Non-perturbative QCD is a good candidate to
answer some of these questions. Phenomenological studies are also very
essential. \\ \\ 

\noindent
{\bf Acknowledgment} \\
This work has been supported by the Japanese Society for the Promotion of
Science. I would like to thank A.~Hosaka, A.~Nakamura and H.~Toki for
fruitful discussions.

\end{document}